# Tuning of crystal structure and magnetic properties by exceptionally large epitaxial strains

J. Buschbeck[1,2], I. Opahle[1,3], M. Richter[1], U.K. Rößler[1], P. Klaer[4], M. Kallmayer[4], H. J. Elmers[4], G. Jakob[4], L. Schultz[1,2], S. Fähler[1]

[1] *IFW Dresden, P.O. Box 270116, 01171 Dresden, Germany*

[2] *Department of Mechanical Engineering, Institute for Materials Science, Dresden University of Technology, 01062 Dresden, Germany*

[3] *Institut für Theoretische Physik, Universität Frankfurt, 60438 Frankfurt/Main, Germany*

[4] *Institut für Physik, Johannes Gutenberg-Universität Mainz, 55099 Mainz, Germany*

**Abstract**

Huge deformations of the crystal lattice can be achieved in materials with inherent structural instability by epitaxial straining. By coherent growth on seven different substrates the in-plane lattice constants of 50 nm thick $Fe_{70}Pd_{30}$ films are continuously varied. The maximum epitaxial strain reaches 8.3 % relative to the fcc lattice. The in-plane lattice strain results in a remarkable tetragonal distortion ranging from $c/a_{bct}$ = 1.09 to 1.39, covering most of the Bain transformation path from *fcc* to *bcc* crystal structure. This has dramatic consequences for the magnetic key properties. Magnetometry and X-ray circular dichroism (XMCD) measurements show that Curie temperature, orbital magnetic moment, and magnetocrystalline anisotropy are tuned over broad ranges.

Strain effects on functional materials are of great current interest for improving materials properties. By strained epitaxial film growth, physical properties can be controlled and improved e.g. in semiconductors [1], multiferroic materials [2] and ferromagnets [3-7]. Moreover, there are even materials exhibiting their functional properties like ferroelectricity only in strained films [8].

During epitaxial growth, the film orientation is controlled by the substrate onto which the material is deposited. In addition, a thin film may also adapt its in-plane lattice parameters to the substrate, even if their equilibrium lattice parameters differ considerably. The particular case when the lattice parameters of the film material are strained such that they are identical to those of the substrate is called strained coherent film growth. In common rigid metals, straining of the crystal lattice by coherent film growth requires substantial elastic energy. As a consequence, already at strains of a few percent, coherent growth is limited to ultrathin films with thicknesses of up to several atomic layers. For the growth of coherent epitaxial films with high strains and large thickness soft materials should be used, as suggested by van der Merwe [9]. Exceptional softening is observed in crystals with lattice instabilities, e.g. in materials with a martensitic transformation. As an example how to exploit such a martensitic instability, Godlevsky and Rabe [10] predicted the possibility to induce a cubic to tetragonal distortion with $c/a_{bct}$ ratios from 0.95 to 1.18 in the magnetic shape memory material $Ni_2MnGa$. In fact, in experiments Dong et al. [11] demonstrated a considerable epitaxial strain of 3% in a Ni-Mn-Ga film.

This shows that for improved tunability of crystal lattice and functional properties in thicker films it is advantageous to exploit lattice instabilities. Here, we report the preparation of seven evenly distributed stages along the Bain path, between face centered cubic *fcc* and body centered cubic *bcc* structure in 50 nm thick $Fe_{70}Pd_{30}$ films (Fig. 1). The martensitic and ferromagnetic $Fe_{70}Pd_{30}$ alloy is well known for anomalous

4magnetomechanical effects such as the magnetic shape memory effect[12]. We demonstrate the inverse magnetomechanical behavior of this compound: By varying $c/a_{bct}$, Curie temperature, orbital magnetic moment, and magnetocrystalline anisotropy are tuned in wide ranges. The approach can be extended to a multitude of materials with ferroelastic or martensitic lattice instabilities enabling exceptionally large strains. In particular, the possibility to stabilize intermediate lattice geometries and vary smoothly between stable phases along Bain transformation paths offers new routes to adjust and understand structure-property relations in functional materials.

Fe-Pd has been suggested as a model system for martensitic transformations, already in 1938 [13]. For the present experiments it is in particular favorable that the $Fe_{70}Pd_{30}$ alloy exhibits a structural instability resulting in softening of the crystal lattice near room temperature[14]. In addition, growth at ambient temperature frequently enables preparation of coherent films with larger thickness due to the reduced mobility of dislocations [15]. In the composition range from $Fe_{70}Pd_{30}$ to $Fe_{75}Pd_{25}$ and in the vicinity of room temperature four phases with different tetragonal distortions have been observed in quenched, chemically disordered bulk: *fcc*, "*fct*", *bct* and *bcc* [16]. These structures follow the Bain transformation path[17]. The Bain path is a geometrical description for the transformation from *fcc* to *bcc* lattice. Considering a body centered tetragonal (*bct*) unit cell with the lattice parameters $a_{bct}$ and $c_{bct}$ (marked by colored atoms and thicker lines in Fig. 1B) one can describe the transformation from *fcc* (top) to *bcc* lattice (bottom) by a continuous variation of the tetragonal distortion $c/a_{bct}$ from $\sqrt{2}$ to 1. In accordance with Bain we use the *bct* description of the unit cells here. The four phases are described by $c/a_{bct}$ ratios of *1.41* (*fcc*), *1.33* ("*fct*"), *1.02* (*bct*) and 1 (*bcc*).

$Fe_{70}Pd_{30}$ films of 50 nm nominal thickness were deposited onto different substrates at room temperature by magnetron sputtering from a 2 inch $Fe_{70}Pd_{30}$ alloy target. The substrates consist of a MgO(100) single crystal with different epitaxial metallic buffer layers (Rh, Ir, Pt, Pd, Au-Cu, Fe and Cr). Details on the growth of the epitaxial films are described in the auxiliary material available online[18]. The epitaxial growth on the different substrates induces considerable distortion of the crystal lattice. X-ray diffraction (XRD) was measured at room temperature. As an overview over the crystal structure, *θ-2θ* diffraction patterns were recorded in Bragg-Brentano geometry (Co K$_\alpha$, 0.1789 nm). The diffraction patterns show a (002)$_{bct}$ reflection of the epitaxial Fe$_{70}$Pd$_{30}$ film (Fig. 1A, reflection marked by an arrow). When varying the substrate this reflection shifts by 12.3 degrees. However, it always lies within the boundaries of the Bain-transformation, between the reflection positions expected for *fcc* and *bcc* structure. Intensity measured at *2θ* ≤ 56° is due to the (002) reflection of the *fcc* Au-Cu, Pt, Pd, Ir or Rh layer.

Determination of the lattice constants was performed in a 4-circle set-up with Euler stage (Cu K$_\alpha$, 0.1540 nm) [18]. The in-plane lattice parameters of the films are identical to the substrate lattice spacings (Fig. 1C). This proves a coherent epitaxial growth in spite of a large variation of the substrate lattice spacing. The epitaxial strain reaches a maximum value of 8.3 % relative to the Fe$_{70}$Pd$_{30}$ fcc lattice [19]. Due to the coherent growth, our films are in a single variant state with both in-plane lattice parameters $a_{bct}$ of the crystal structure's basal plane fixed by the cubic substrate lattice, while no constraint exists for the lattice parameter $c_{bct}$ in the out-of-plane direction. The measured $c/a_{bct}$ ratio ranges from $c/a_{bct}$ = 1.09 to 1.39, covering most of the Bain path from *bcc* to *fcc*. This corresponds to a variation of tetragonal distortion by 27% which we could stabilize in films with a remarkable thickness of 50 nm. The dotted line in Fig. 1C represents the ideal behavior at constant unit cell volume ($c/a_{bct} = V_{bct}/2a^3$, $V_{bct}$ = 0.0265 nm$^3$ [19]). The experimental data points show that the assumption of constant volume is justified even for such large deformations.

The achieved strained film growth indicates very low energy differences between the crystal structures along the Bain path. For the present material, this is confirmed by density functional calculations (Fig. 2A). Electronic structure calculations for disordered Fe-Pd alloys were performed with the full potential local orbital (FPLO) code [20] in the framework of density functional theory. Disorder was treated in the coherent potential approximation (CPA) [21] and the exchange and correlation potential was treated in the local spin density approximation (LSDA). In the calculations, the $c/a_{bct}$ ratio is varied, while the volume of the unit cell is held constant, as justified by our experiments. Numerical details of the calculations are identical to those given in Ref. 22. In comparison to the rigid pure Fe [23] and Pd [24], the energy landscape is very flat along the entire Bain path, i.e., energy differences between the different $c/a_{bct}$ ratios are indeed small. Hence, the film-substrate interaction is sufficient to stabilize large tetragonal distortions in films of bulk-like thickness.

The key property of magnetic shape memory alloys is their magnetocrystalline anisotropy. Strained epitaxial growth holds the promise of adjusting this property, beyond what is possible by conventional inverse magnetostriction. On the atomic scale, the origin of the magnetocrystalline anisotropy is the spin-orbit coupling of the valence electrons in combination with exchange-splitting and with the particular electronic structure. We studied the effect of lattice distortion on the spin-orbit related properties by measuring the ratio of orbital- to spin magnetic momentum ($\mu_{orb}/\mu_{spin}$) of the iron 3d electrons by means of X-ray magnetic dichroism along the [001] crystal direction of four films with considerably different tetragonal distortion. In addition we determined the anisotropy constants $K_1$ and $K_3$ of these films by magnetometer measurements along different crystallographic directions [001]; [100]; [110] (Fig. 2 B). Magnetic measurements with both methods were performed at 300K [18]. While $K_1$ defines the work required to magnetize the material along the hard magnetization axis [001], $K_3$ is a measure of the anisotropy in the basal plane of the tetragonal unit cell. All three quantities ($\mu_{orb}/\mu_{spin}$ and $K_1$, $K_3$) show considerable changes along the Bain path but similar trends. At $c/a_{bct}$ = 1.33 maximum anisotropy is observed. This value of tetragonal distortion coincides with the *"fct"* phase that exhibits high magnetocrystalline anisotropy and shows the magnetic shape memory effect in bulk. For the magnetic shape memory effect, the anisotropy constant $K_1$ is a crucial quantity. We find that due to the strain and tetragonal distortion $K_1$ is strongly varying in our series of epitaxial films. When approaching cubic structures the magnetic anisotropy is reduced as expected due to their high symmetry. The maximum value of $K_1 = 1.5*10^5$ J/m$^3$ at $c/a_{bct}$=1.33 is identical with the literature value that has been reported at $T$=100 K for a *"fct"* single crystal with $c/a_{fct}$ = 0.94 (corresponding to $c/a_{bct}$ = 1.34) [25].

Moreover, the tetragonal distortion of the lattice also significantly changes the Curie temperature. Temperature dependent spontaneous magnetization was measured in the film plane (Fig. 3). The applied field of 1 T is sufficient to saturate the sample along this direction. Increase in curvature is observed with increasing $c/a_{bct}$ ratio. To explain this behavior, the Curie temperature $T_C$ was evaluated. According to Kuz'min's model [26], $T_C$ can be determined by a fit of the relative magnetization curves (inset in Fig. 3). The additional shape parameter $s$ that is included in this model, depends on the spin wave stiffness[27]. While the shape parameter does not vary significantly with tetragonal distortion, the ferromagnetic transition temperature increases remarkably from 652 K at $c/a_{bct}$ = 1.39 to a value of 829 K at $c/a_{bct}$ = 1.09. The extrapolated value of $T_C$ = 650 K at $c/a_{bct}$ = 1.41 (*fcc*) is similar to the literature value of 600 K reported for the fcc-phase in Fe$_{70}$Pd$_{30}$ bulk [28, 29].

In conclusion, by strained coherent growth of 50 nm thick Fe$_{70}$Pd$_{30}$ films with inherent structural instability, we achieve a quasi continuous variation of the lattice distortion along the Bain path. The magnetic properties of the Fe$_{70}$Pd$_{30}$ films display large changes: The Curie temperature is increased more than 25% with respect to the value for $Fe_{70}Pd_{30}$ with fcc structure. The ratio of orbital vs. spin magnetic momentum changes by a factor of two. This is accompanied by a large modification of the magnetic anisotropy from near zero to values close to those of "*fct*" bulk $Fe_{70}Pd_{30}$. Softening of the crystal lattice and a flat energy landscape along the Bain path are not a unique feature of this alloy. Similar lattice instabilities may be exploited in various functional materials including (magnetic) shape memory, ferroelectric, multiferroic, or magnetocaloric materials for extended adjustability of their crystal structure in strained epitaxial films. This must cause severe changes in the electronic structure of the materials and, thus, enables to wider tune their magnetic, transport, optical, or even catalytic properties.

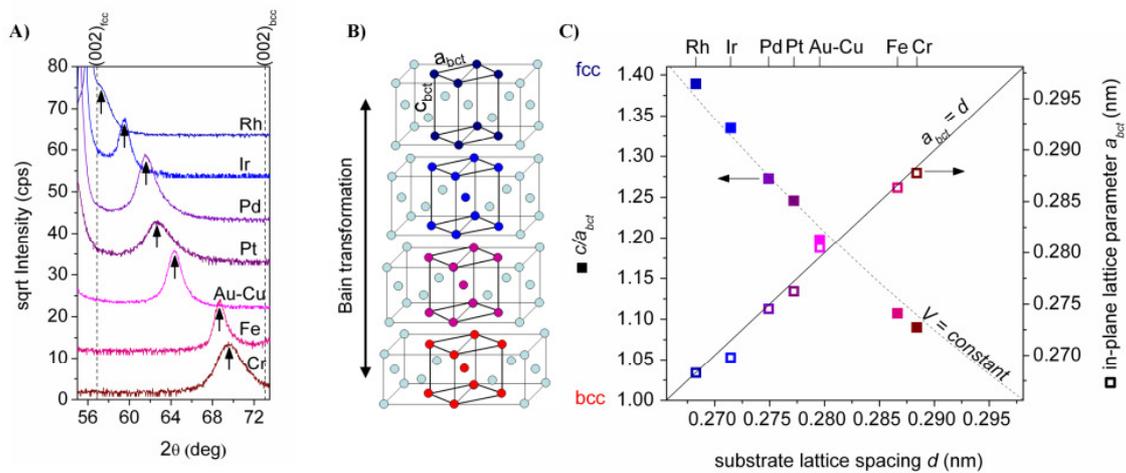

FIG. 1 : **A:** A considerable shift of the $Fe_{70}Pd_{30}$ $(002)_{bct}$ peak (arrow) observed in X-ray diffraction patterns is measured on the films grown on different substrate materials ($CoK_\alpha$). Peak positions expected for *fcc* and *bcc* structure are marked by dotted lines. They represent the boundaries of the Bain transformation. **B:** Sketch of the different stages during the Bain transformation between *fcc* (top, dark blue) and *bcc* structure (bottom, red). The black lines mark the *bct* unit cell used to describe this transformation. **C:** Variation of the $Fe_{70}Pd_{30}$ crystal lattice (substrate materials are marked on top). The in-plane lattice parameter $a_{bct}$ (open symbols) follows the straight line representing identity with the substrate's lattice spacing *d*. The $c/a_{bct}$ ratio (solid symbols) is varied almost from *fcc* to *bcc* structure by the coherent epitaxial growth. The dotted curve illustrates the expected change of $c/a_{bct}$ at constant volume of the unit cell.

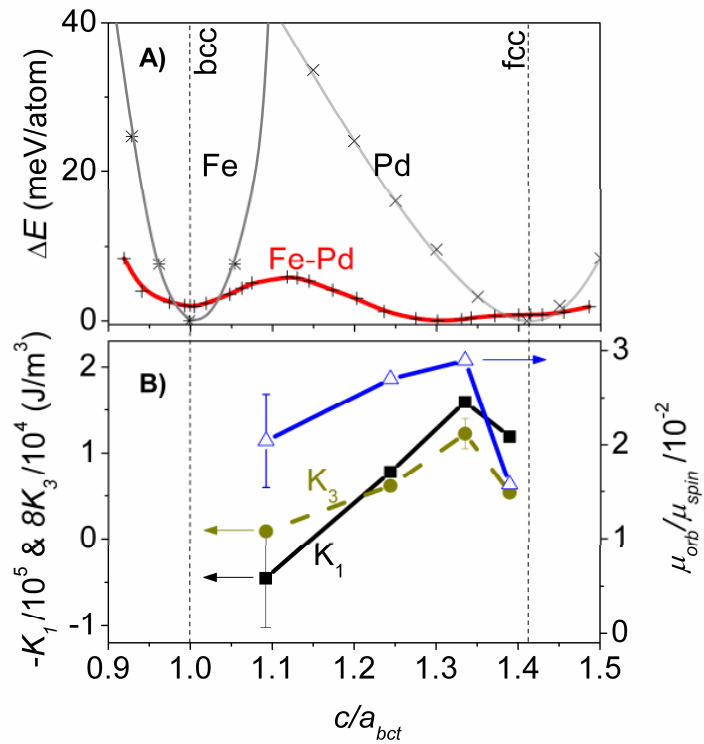

FIG. 2: **A**: Calculated energy required to induce tetragonal distortions ($c/a_{bct}$) along the Bain path. Compared to literature data of the common metals Fe [23] and Pd [24] only little elastic energy is required for a tetragonal distortion of Fe-Pd. **B:** Change of the magnetocrystalline anisotropy constants $K_1$ and $K_3$ and the ratio of orbital to spin momentum ($\mu_{orb}/\mu_{spin}$) of Fe 3d electrons along the Bain path. For each quantity the experimental error is represented by an error bar.

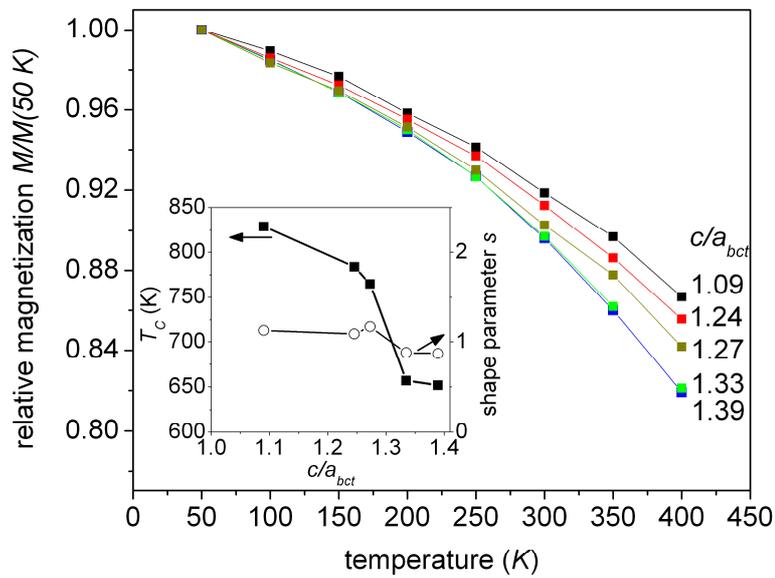

FIG. 3: The curvature of the relative spontaneous magnetization versus temperature decreases with increasing tetragonal distortion. **Inset:** Curie temperature $T_C$ and shape parameter s that have been extracted using Kuz'min's model [26]. $T_C$ increases considerably when approaching the *bcc* structure. The shape parameter only shows minor changes.

**Acknowledgements** The authors thank O. Heczko, M. E. Gruner, J. McCord and S. Kaufmann for helpful discussions and U. Besold and T. Eichhorn for experimental support. This work is funded by DFG via the priority program on magnetic shape memory alloys (www.magneticshape.de).

**References**

[1] J. C. Bean, Science **230**, 127 (1985).

[2] J. Wang, J. B. Neaton, H. Zheng, V. Nagarajan, S. B. Ogale, B. Liu, D. Viehland, V. Vaithyanathan, D. G. Schlom, U. V. Waghmare, N. A. Spaldin, K. M. Rabe, M. Wuttig, R. Ramesh, Science **299**, 1719 (2003).

[3] C. Thiele, K Dörr, S. Fähler, L. Schultz, D. C. Meyer, A. A. Levin, P. Paufler, Appl. Phys. Lett. **87**, 262502 (2005).

[4] A. Winkelmann, M. Przybylski, F. Luo, Y. Shi, J. Barthel, Phys. Rev. Lett. **96**, 257205 (2006).

[5] X. W. Li, A. Gupta, Xiao Giang, Appl. Phys. Lett. **75**, 713 (1999) .

[6] M. J. Pechan, C. Yua, D. Carr, C. J. Palmstrøm, J. Magn. Magn. Mat. **286**, 340 (2005).

[7] A. R. Kwon, V. Neu, V. Matias, J. Hänisch, R. Hühne, J. Freudenberger, B. Holzapfel, L Schultz, S. Fähler, New J. Phys. **11**, 083013 (2009)

[8] J. H. Haeni, P. Irvin, W. Chang, R. Uecker, P. Reiche, Y. L. Li, S. Choudhury, W. Tian, M. E. Hawley, B. Craigo, A. K. Tagantsev, X. Q. Pan, S. K. Streiffer, L. Q. Chen, S. W. Kirchoefer, J. Levy, D. G. Schlom, Nature **430**, 758 (2004).

[9] J. H. van der Merwe, J. Appl. Phys. **34**, 123 (1963).

[10] V. V. Godlevsky, K. M. Rabe, Phys. Rev. B **63**, 134407 (2001).

[11] J. W. Dong, J. Lu, J. Q. Xie, L. C. Chen, R. D. James, S. McKernan, C. J. Palmstrøm, Physica E **10**, 428 (2001).

[12] R. D. James, M. Wuttig, Philos. Mag. A **77**, 1273 (1998).

[13] R. Hultgren, C. A. Zapffe, Nature **142**, 395 (1938).

[14] R. Oshima, S. Muto, F. E. Fujita, Mat. Trans. JIM **33**, 197 (1992).

[15] R. Hull, J. C. Bean, Crit. Rev. Solid State Mater. Sci. **17**, 507 (1992).

[16] M. Sugiyama, R. Oshima, F. E. Fujita, Trans. Jpn. Inst. Met. **25**, 585 (1984).

[17] E. C. Bain, Trans. Am. Inst. Min. Met. Eng. **70**, 25 (1924)

[18] See EPAPS Document No. [number will be inserted by publisher] for more details on experimental methods and the film texture. For more information on EPAPS, see http://www.aip.org/pubservs/epaps.html.

[19] J. Cui, T. W. Shield, R. D. James, Acta Mater. **52**, 35 (2004).

[20] K. Koepernik, H. Eschrig, Phys. Rev. B **59**, 1743-1757 (1999); http://www.FPLO.de.

[21] K. Koepernik, B. Velicky, R. Hayn, H. Eschrig, Phys. Rev. B **55**, 5717 (1997).

[22] I. Opahle, K. Koepernik, U. Nitzsche, M. Richter, Appl. Phys. Lett. **94**, 072508 (2009).

[23] S. L. Qiu, P. M. Marcus, H. Ma, Phys. Rev. B, **64**, 104431 (2001).

[24] F. Jona, P. M. Marcus, Phys. Rev. B **65**, 155403 (2002).

[25] T. Kakeshita, T. Fukuda, T. Takeuchi, Mat. Sci. Eng. A **438-440**, 12 (2006).

[26] M. D. Kuz'min, Phys. Rev. Lett. **94,** 107204 (2005).

[27] M. D. Kuz'min, M. Richter, A. N. Yaresko, Phys. Rev. B **94**, 107204 (2006).

[28] A. Kussmann, K. Jessen, J. Phys. Soc. Jpn. **17**, 136 (1962).

[29] M. Matsui and T. Shimizu and H. Yamada and K. Adachi, J. Magn. Magn. Mat. **15**, 1201 (1980).

# Auxiliary Material

Here we provide experimental details of the film deposition as well as the methods for structural and magnetic characterization and density functional calculations. This is followed by figures supporting the results of the structural characterization presented in the main paper. Film texture is characterized by means of pole figure measurements. By these, epitaxial film growth is proven and the orientation relationships of film, buffer and substrate are determined. In addition, the positions of the poles in the pole figure verify the tetragonal distortion in an independent measurement. Finally results of the reciprocal space mapping are presented which are used to exemplarily verify the coherent epitaxial growth of a $Fe_{70}Pd_{30}$ film deposited on Pt buffer.

In the main paper, for better readability, the term "substrate" describes the metals where the $Fe_{70}Pd_{30}$ films were deposited on. These metals themselves are prepared as epitaxial thin layers on MgO(100) single crystals. Thus, for precise designation of the parts of these composite substrates, we will name this metal layer "buffer" in the following.

**Methods**

*Film deposition*

Using the sputtering method for $Fe_{70}Pd_{30}$ film deposition we could overcome the limitations of our previous experiments where pulsed laser deposition (PLD) was applied [26]. For PLD films we observed that in addition to the intended stabilization of *fct* martensite severe stress of up to 4 GPa caused (111) deformation twinning in the film, resulting in the loss of epitaxy at film thickness exceeding 20 nm. We could attribute the origin of the stress to high kinetic energy of the deposited ions characteristic for PLD [27]. To reduce film stress, we changed to the sputtering method for deposition of $Fe_{70}Pd_{30}$ films where the kinetic energy of the ions is at least one order of magnitude lower compared to PLD [28].

The base pressure in the sputtering chamber was less than $10^{-8}$ mbar and Argon sputtering gas of *6N* purity was used. Epitaxial metallic buffer layers (Rh, Ir, Pt, Pd, Au-Cu, Fe and Cr) serving as substrates for the strained growth of $Fe_{70}Pd_{30}$ films were deposited at 300°C onto MgO(100) single crystals of 10 mm x 10 mm x 0.15 mm in size. Buffer layers (Cr, Fe, Au-Cu, Pd, Pt) were prepared using magnetron sputtering from 2 inch targets. In order to limit materials cost, Rh and Ir buffers were prepared using on-axis pulsed laser deposition from smaller targets and subsequently transferred ex-situ to the sputtering chamber. The nominal film architecture is: MgO / 50 nm buffer / 50 nm $Fe_{70}Pd_{30}$. When metals with *fcc* crystal structure are used as buffer (Rh, Ir, Pt, Pd or Au-Cu), prior deposition, a thin layer of 5 nm Cr was grown on MgO to promote epitaxial growth. Buffer layer of metals with bcc-structure (Cr and Fe) grow epitaxial without a further interlayer.

To promote subsequent $Fe_{70}Pd_{30}$ film growth on a relaxed buffer surface, the thickness of the buffer layer was chosen such that it exceeds the common critical thickness of metal layers by more than one order of magnitude. Deposition at 300°C enables epitaxial growth of the buffer layers with bulk-like lattice parameters. By depositing $Fe_{70}Pd_{30}$ at room temperature undesirable decomposition into ordered $L1_0$ phase and iron rich Fe(Pd) phase is avoided [29].

Film composition was checked for films on Rh, Ir, Pt and Cr substrate by energy dispersive x-ray spectroscopy using a bulk $Fe_{70}Pd_{30}$ standard. The measured film composition of $Fe_{70.4}Pd_{29.6}$ (±0.7) is constant and identical to the target composition.

*Structural Characterization*

For diffraction pattern measurements in Bragg-Brentano geometry a 2-circle Phillips set-up is used (CoKα). Since the *bcc* lattice normally does not exist at a composition of $Fe_{70}Pd_{30}$ under normal conditions, the lattice spacing of the *bcc* structure was calculated assuming constant volume by $V_{bcc}=V_{fcc}/2$ = 0.0265 nm$^3$ [30]. For determination of the crystal structure of the $Fe_{70}Pd_{30}$ films, as shown in Fig. 1C of the main paper, *2θ* of $(002)_{bct}$ and *2θ* of $(101)_{bct}$-type lattice planes were measured in 4-circle geometry. By this two independent, non-collinear lattice parameters are probed. Since both in-plane lattice parameters are identical this is sufficient to describe the tetragonal unit cell. Using elementary geometry we calculated the out-of-plane lattice constant $c_{bct}$ and the in-plane lattice constants $a_{bct}$ by $c_{bct} = 2 \cdot d(002)_{bct}$ and $a_{bct} = \sqrt{\dfrac{d(101)_{bct}^2 \cdot c_{bct}^2}{c_{bct}^2 - d(101)_{bct}^2}}$.

From these values the $c/a_{bct}$ ratio was calculated. For $Fe_{70}Pd_{30}$ grown on Rh buffer, the $(101)_{bct}$ and $(111)_{fcc}$ lattice planes of epitaxial $Fe_{70}Pd_{30}$ film and Rh buffer, respectively, exhibit very similar lattice spacing and spatial alignment. Since film and buffer could not be distinguished in the Phillips 4-circle device in this case, the lattice parameters of this film were measured in the reciprocal space mapping setup having a higher resolution. Reciprocal space measurements were performed in a non-commercial 4- circle set-up using a rotating anode system (*Cu K$_\alpha$*), equipped with a focussing multilayer X-ray mirror. Besides conventional angular scans, the software allows direct mapping of scattered intensity in reciprocal space coordinates.

*Magnetic Characterization*

X-ray magnetic circular dichroism (XMCD) was measured in transmission at the German synchrotron light source BESSY II (beam line UE56/1-SGM) at normal incidence. XMCD results are obtained from transmission measurements detecting the X-ray intensity transmitted through the $Fe_{70}Pd_{30}$ films via the luminescence light the X-ray light induces in the MgO(100) substrate [31],[32]. This transmission data measures film properties averaged along the film normal. The linear background subtraction was adjusted in order to achieve the same relative post-edge absorption intensity at 750 eV as for the transmission data. The magnetization of the sample was switched at each energy step by applying a magnetic field of 1.22 T at opposite directions perpendicular to the sample. Measurements were performed at 300 K.

Magnetic anisotropy was characterized by hysteresis curve measurements at 300 K up to fields of 4.5 T using a Quantum Design Physical Property Measurement System (PPMS) with vibrating sample magnetometer (VSM) add-on.

The magnetocrystalline anisotropy energy of a tetragonal unit cell can be described by the following equation [33]:

$$E_A = K_1 \sin^2 \phi + K_2 \sin^4 \phi + K_3 \sin^4 \phi \cdot \cos 4\alpha \qquad [1]$$

where $K_1...K_3$ are the anisotropy constants; $\phi$ is the angle between magnetization direction and c-axis and $\alpha$ is the angle to the a-axes in the basal plane of the tetragonal unit cell. It is known from literature that in $Fe_{70}Pd_{30}$ *"fct"* bulk single crystals the hard magnetization direction is aligned along the c-axis while the a-axes are the easy magnetization directions [30].

Finite values of $K_1$ cause a linear dependence of the magnetization on the applied magnetic field along the hard magnetization axis. Finite values of the second anisotropy constant $K_2$ cause a deviation from this linear dependence. Since we observe almost linear behavior we assumed $K_2 \approx 0$. The 4-fold anisotropy in the basal plane of the tetragonal unit cell is described by $K_3$.

$K_1$ was determined from the anisotropy field obtained by linear extrapolation of hysteresis loop measurements along the [001] crystal direction of the film by $K_1 = -H_A/2J_S$. Shape anisotropy was corrected by assuming a demagnetization factor of $N=1$ of an infinite film. $K_3$ was determined from the area $A$ enclosed by the hysteresis loops measured along the [100] and [110] directions according to $2K_3 = W[100]_{bct} - W[110]_{bct} = A$ with W[hkl]$_{bct}$ being the work to magnetize the sample along the respective direction of the bct unit cell. $K_3$ of a *"fct"* single crystal at different temperatures and applied stresses has been reported by Cui et al. [30]. Accordingly, in *"fct"*, $a$ and $b$ axes of the unit cell are the easy magnetization directions. In agreement with the Bain transformation we observe the [110] direction being the easy axis in *bct* description. However, due to the different descriptions of the anisotropy energy and unit cell used in the work of Cui et al and here, the absolute values of $K_3$ have to be converted. A factor of 1/8 has to be applied to the values published by Cui et al. [30] in order to make them comparable [34]. Cui et al. report $K_{3(Cui)}/4 \approx 1.2*10^4$ J/m$^3$ at $c/a_{fct} = 0.935$ (-20°C) under applied stresses of 2 and 8 MPa. Due to the applied stress, the single crystal was in the single variant state. In *bct* description $c/a_{fct} = 0.93$ corresponds to $c/a_{bct} = 1.32$. When considering the different measurement temperatures of 253 K and 300 K, respectively, the values of $1/8K_{3(Cui)}=K_3$ $=6*10^3$ J/m$^3$ and $K_3 =1.3*10^3$ J/m$^3$ measured here for $c/a_{bct}$=1.33, are in reasonable agreement.

Spontaneous magnetization $J_S$ was determined at a field of 1 T from the demagnetizing branch of the in-plane hysteresis measurements, after magnetizing the sample in a magnetic field of 2 T. Film thickness was determined by X-ray reflectometry. In-plane, the samples saturate at fields below 0.5 T.

*Energy calculation along the Bain path*

Experimentally, the martensitic transformation from *fcc* ($c/a_{bct}$=1.41) to *"fct"* ($c/a_{bct}$≈1.34) in disordered Fe-Pd alloys is found for compositions where the ground state changes from *fcc* to *bcc* with increasing Fe content. In the vicinity of this transition, the energy curve along the Bain path becomes flat. In the LSDA calculations of the internal energy this transition is found at somewhat higher Fe concentrations between 78 and 82 at.% in the vicinity of the experimental volume, compared to the experimental concentration of 70 at.% Fe. The deviation between the experimental phase diagram and the calculations for the transition between the *fcc* and *bcc* structures can be mainly attributed to the entropy contributions in the thermodynamic potential at finite temperatures that are not included in the electron-theoretical calculation at zero temperature. Fig.1 shows the calculated total energy along the Bain path for Fe$_{78}$Pd$_{22}$ with the atomic volume fixed to 0.013256 nm$^3$. The specific shape of the energy curve as function of the distortion along the Bain path depends on the chosen composition, but the very flat energy landscape is characteristic for alloys with an *"fct"* martensite phase.

**Auxiliary Structural Characterization**

*Texture measurements*

Epitaxial film growth on the buffer is proven by the existence of a single pole in each quadrant of the Fe$_{70}$Pd$_{30}$ (101)$_{bct}$ pole figure measurements (Fig. S2). The four-fold symmetry of the used (001) oriented MgO single crystal surface makes it sufficient to depict just a quarter of the pole figure.

Orientation relationships of film‖buffer‖MgO(001) have been determined as follows:

Fe$_{70}$Pd$_{30}$*bct*(001)[110]‖*fcc*-buffer(001)[100]‖Cr(001)[110]‖MgO(001)[100] or

Fe$_{70}$Pd$_{30}$*bct*(001)[110]‖*bcc*-buffer(001)[110]‖ MgO(001)[100]

for the films on *fcc*-buffers (Rh, Ir, Pt, Pd, Au-Cu) and *bcc*-buffer (Fe,Cr), respectively. The orientation relationships are sketched in Fig. S3. The *bct* unit cell of the Fe$_{70}$Pd$_{30}$ film is 45° rotated with respect to the MgO fcc unit cell independent of the buffer layer. In contrast, the relative orientation of the *bct* unit cell of the Fe$_{70}$Pd$_{30}$ film with respect to the buffer, changes from [100]$_{bct}$(film)‖ [110]$_{fcc}$ on *fcc*-buffer to [100]$_{bct}$(film)‖[100]$_{bcc}$ on *bcc*-buffer. However, the epitaxial orientation relationship of film and buffer remains identical when the buffers crystal structure is described according to the Bain concept and a *bct* unit cell with *c/a*=1.41 is used for *fcc*-buffers (not shown in the sketch). When doing so, the orientation relationship simplifies to [100]$_{bct}$(film) ‖ [100]$_{bct}$(buffer). Thus, as done in the paper, we can use the "substrate lattice spacing d", equal to d[100]$_{bct}$(buffer), as the key parameter for coherent epitaxial growth, independent of the actual crystal structure of the used buffers. In the common *fcc* and *bcc* description the lattice spacing [100]$_{bct}$ is equal to [220]$_{fcc}$ and [100]$_{bcc}$. At $c/a_{bct}$=1.41 (*fcc*) the (101)$_{bct}$ pole is identical to the (111)$_{fcc}$ pole.

The four-fold symmetry of the buffer forces both in-plane lattice constants to be identical, resulting in the observed tetragonal distortion. In addition to texture and the epitaxial relationship, pole figures also give information about the crystal structure of the film itself. The position of the (101)$_{bct}$ pole along the tilt angle $\Psi$ is a measure of the tetragonal distortion of the *bct* unit cell. As sketched in Fig. S2, $\Psi$ corresponds to the angle between the (101)$_{bct}$ lattice plane and the substrate plane. When the structure changes from $c/a_{bct}$=1.41 (*fcc*) to $c/a_{bct}$=1 (*bcc*), a position change of the (101)$_{bct}$ poles from $\Psi = 54.7°$ to $\Psi = 45°$ is expected. Indeed, this shift of position is already visible in exemplarily shown pole figures in Fig. S2. The measured angle $\Psi$ for Fe$_{70}$Pd$_{30}$ on different buffers is summarized in Fig. S4. When varying the buffer lattice spacing, a continuous shift of the (101)$_{bct}$ poles is observed, between the $\Psi$ angles expected for *fcc* and *bcc* structure. The variation of the (101)$_{bct}$ pole positions shows that, following the Bain path, *bct* crystal structures have been stabilized with intermediate tetragonal distortions between $c/a_{bct}$=1.41 and 1. As an independent measurement, this behaviour confirms the previously depicted dependence of the $c/a_{bct}$-ratio versus substrate lattice spacing depicted in Fig. 1B.

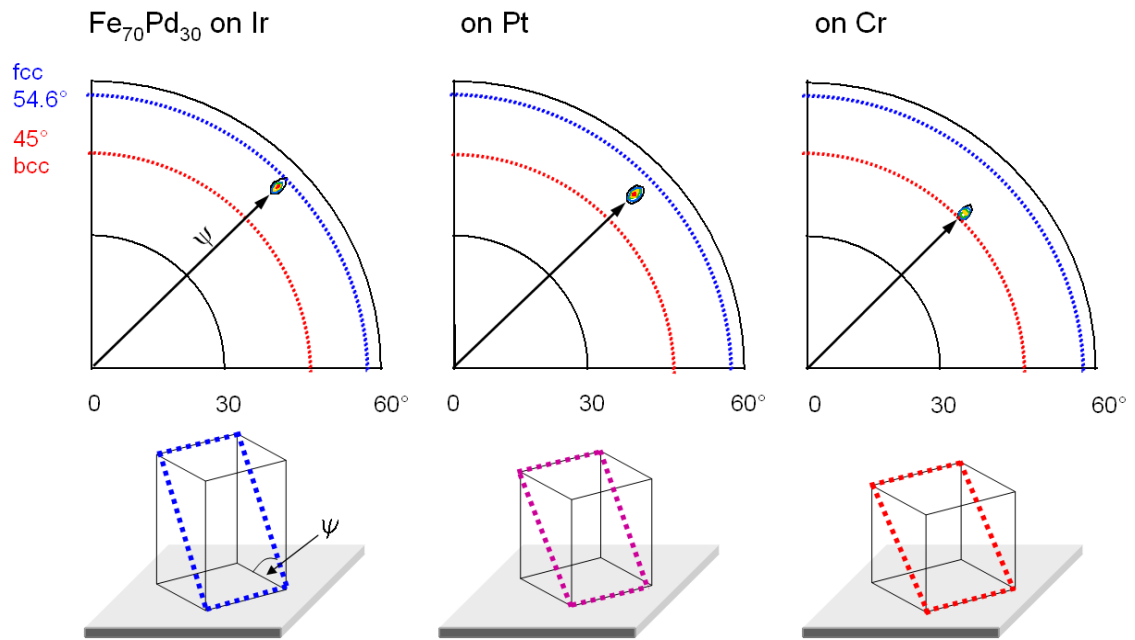

Figure S1 : $(101)_{bct}$ pole figures measured on $Fe_{70}Pd_{30}$ films deposited on Ir, Pt and Cr buffers verify the epitaxial growth. A variation of the tetragonal distortion causes a variation of the tilt angle $\Psi$ of the $(101)_{bct}$ lattice planes (sketch). In result, a shift of the pole position along $\Psi$ is measured in the pole figure. Blue and red dotted lines within the pole figures mark the tilt angles $\Psi$ expected for cubic *fcc* or *bcc* structure, respectively.

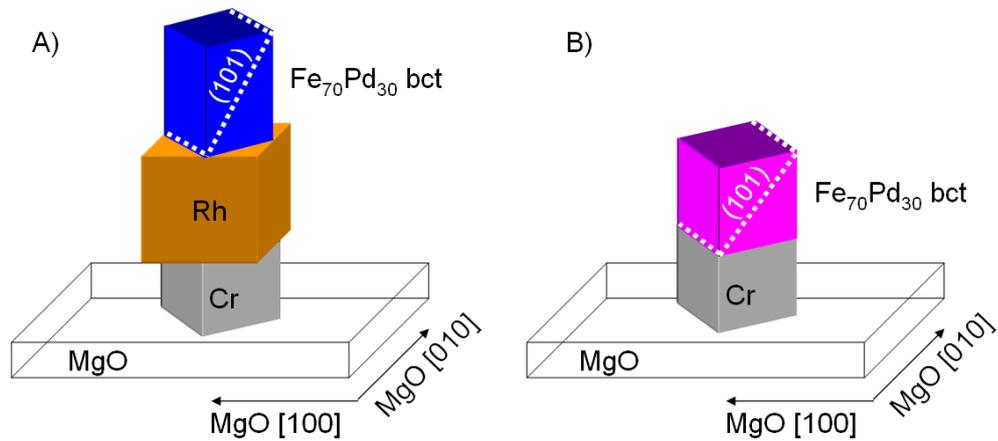

Figure S2: Orientation relationship of the layer systems and the MgO single crystal as determined from pole figure measurements. Two types have to be distinguished: A) $Fe_{70}Pd_{30}$ on *fcc*-buffer using a *bcc* Cr-adhesion layer and B) $Fe_{70}Pd_{30}$ directly deposited on *bcc*-buffer. In the $Fe_{70}Pd_{30}$ unit cell one $(101)_{bct}$ plane is marked by white dotted lines.

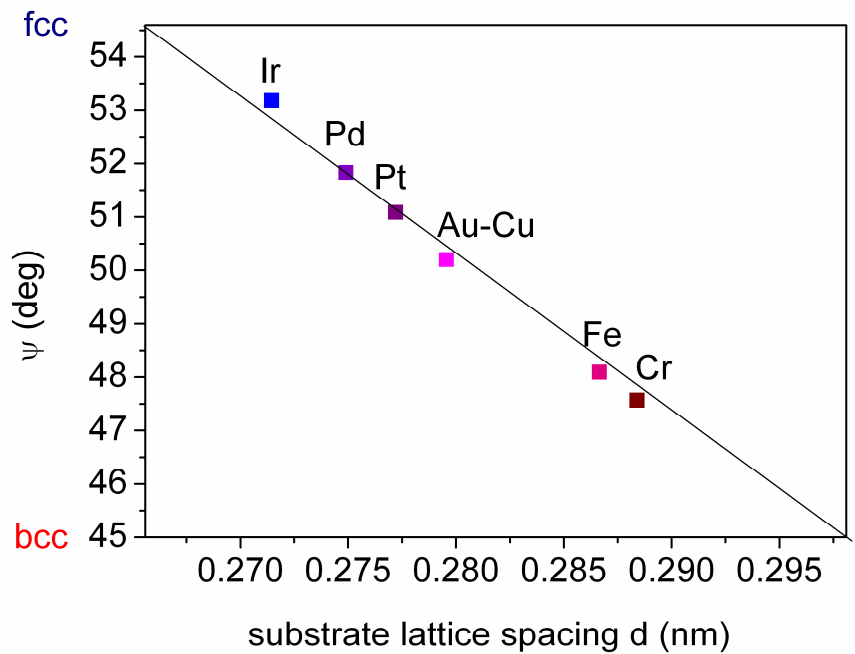

Figure S3: Tilt angle $\Psi$ of the $(101)_{bct}$ reflections extracted from pole figure measurements of $Fe_{70}Pd_{30}$ on different buffers. The behaviour verifies the existence of a tetragonal distortion following the Bain path between *fcc* and *bcc* structure. For the film on Rh, we could not measure $\Psi$ due to the overlapping of film and buffer reflections.

*Reciprocal Space Mapping*

Besides the determination of the crystal structure of the film grown on Rh buffer, we used reciprocal space mapping in order to probe the structure of the $Fe_{70}Pd_{30}$ film on Pt buffer in more detail. This film was chosen since its c/a ratio is approximately in the middle of the Bain path. Figure S5 shows the plane scan of the substrate, buffer and film reflections measured with varying *h* and *l* along a $(101)_{bct}$ plane. The coordinates of the reciprocal space are based on the *bct* unit cell of the film. In this particular measurement plane, all three reflections of $Fe_{70}Pd_{30}$ film $(101)_{bct}$, Pt buffer $(111)_{fcc}$ and $MgO(111)_{fcc}$ are visible. Due to the different lattice spacings of film, buffer and MgO they appear at different positions in reciprocal space. Buffer and film reflections are much broader compared to the $MgO(111)_{fcc}$. In agreement with the less perfect crystal structure expected for the epitaxial films compared to bulk single crystals this can be interpreted as a wider spread of the lattice spacings in the films. Together with further measurements along different measurement planes, the single crystalline state, the orientation as well as the previously calculated tetragonal structure of the $Fe_{70}Pd_{30}$ unit cell could be verified. No evidence for stress relaxation by a modulated crystal structure or lateral splitting into variants (regions) with different *c/a* ratios was observed.

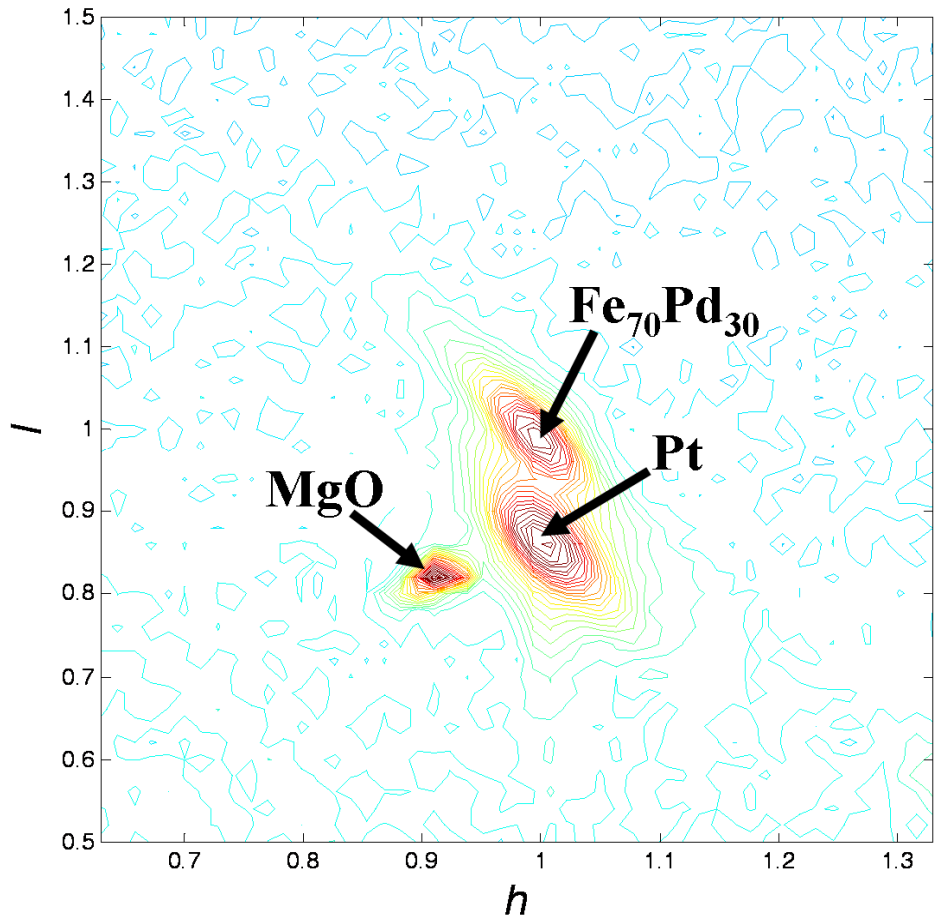

Figure S4: Reciprocal Space Mapping of $Fe_{70}Pd_{30}$ deposited on Pt, showing the $(111)_{fcc}$ reflections of MgO and Pt and a $(101)_{bct}$ reflection of *bct* $Fe_{70}Pd_{30}$ measured by varying *h* and *l* in the reciprocal space coordinate system of the *bct* $Fe_{70}Pd_{30}$. The *h* and *l* coordinate give the in-plane and out-of-plane lattice constants, respectively.

**References**

[26] J. Buschbeck, I. Lindemann, L. Schultz, S. Fähler, Phys. Rev. B **76,** 205421 (2007).

[27] T. Edler, J. Buschbeck, C. Mickel, S. Fähler, S. G. Mayr, New J. Phys. **10**, 063007 (2008).

[28] S. Fähler, H. U. Krebs, Appl. Surf. Sci. **96-98**, 61 (1996).

[29] J. Buschbeck, O. Heczko, A. Ludwig, S. Fähler, L. Schultz, J. Appl. Phys. **103**, 07B334 (2008).

[30] J. Cui, T. W. Shield, R. D. James, Acta Mater. **52**, 35 (2004).

[31] M. Kallmayer, H. Schneider, G. Jakob, H. J. Elmers, K. Kroth, H. C. Kandpal, U. Stumm, S. Cramm, Appl. Phys. Lett. **88**, 072506 (2006).

[32] M. Kallmayer, A. Conca, M. Jourdan, H. Schneider, G. Jakob, B. Balke, A. Gloskovskii, H. J. Elmers, J. Phys. D - Appl. Phys. **40**, 1539 (2007).

[33] K. H. J. Buschow, F. R. de Boer *Physics of Magnetism and Magnetic Materials* (Kluwer, New York 2004) pp. 97-100

[34] Explanation for the origin of the factor of 1/8: The formulation of the anisotropy energy used by Cui et al. is $E_A = K_1 \sin^2\phi + K_2 \sin^4\phi + K_3 \sin^2\alpha \cdot \sin^2\beta$. Compared to our formulation, only the description of the anisotropy in the basal plane differs ($\phi=90°$). In an orthogonal system, instead of using both angles $\alpha$ and $\beta$ to describe the magnetization direction relative to the crystal axes *a* and *b* it is sufficient to consider one of the angles. Hence $\sin^2\alpha \cdot \sin^2\beta$ can be equally expressed by $\sin^2\alpha \cdot \cos^2\alpha = -1/8 \cos 4\alpha$. Thus, if the unit cell remains the same, the conversion factor is -1/8. However, instead of *"fct"* we used the *bct* unit cell here. In the *bct* description of the unit cell, $[100]_{fct}=[110]_{bct}$ while $[110]_{fct}=[100]_{bct}$. Thus the change from *"fct"* description to *bct* description, causes a further change in sign, resulting in a conversion factor of 1/8.